\documentclass[12pt,twoside,a4paper,fleqn]{article}
\usepackage[margin=0.7in,footskip=0.5in]{geometry}

\usepackage{epsfig}
\usepackage{graphicx}
\usepackage{amssymb}
\usepackage{mathrsfs}
\usepackage{dcolumn}

\newcommand{\p}{\prime}

\newcommand{\beq}{\begin{equation}}
\newcommand{\eeq}{\end{equation}}

\newcommand{\ball}{\begin{align}}
\newcommand{\eall}{\end{align}}

\newcommand{\beqar}{\begin{eqnarray}}
\newcommand{\eeqar}{\end{eqnarray}}

\newcommand{\ben}{\begin{enumerate}}
\newcommand{\een}{\end{enumerate}}

\begin{document}

\title{Quantum oscillations of the  Stoner susceptibility : theory}
\author{Ajeet Kumar and Raman Sharma\footnote{Himachal Pradesh University Shimla-171005, India. Email: kumarajeetsuryavansi@gmail.com; bramans70@yahoo.co.in}, Navinder Singh\footnote{Physical Research Laboratory,
Ahmedabad, India, PIN: 380009. Email: navinder.phy@gmail.com; navinder@prl.res.in}}
\maketitle

\begin{abstract}
Oscillatory effects in magnetic susceptibility of free electrons in a strong magnetic field is well
known phenomenon and is well captured by Lifshitz-Kosevich formula. In this paper we point out
similar oscillatory effects in Stoner susceptibility which makes the system to oscillate between
paramagnetic phase and  ferromagnetic phase alternatively as a function of external magnetic field
strength. This effect can happen in a material which is tuned near to its magnetic instability. We 
suggest an experimental set-up to observe this effect. We also suggest that our result can be exploited to control a quantum critical system around its quantum critical point to study its thermodynamical or transport properties.

\vspace{1.5cm}
\end{abstract}
\section{Introduction}
The magnetization of  free electron gas consist of two parts : the first part is due to the
intrinsic magnetic moment(spin) of the electrons called Pauli paramagnetism, and the second part is
due to orbital motion of electrons and is called Landau diamagnetism. These two are weak forms of
magnetism. The possibility of strong form of magnetism i.e. ferromagnetism in a free electron gas
was discussed by Bloch on the basis of exchange interaction between electrons which is
ferromagnetic in nature (parallel spin electrons reduce mutual Coulomb repulsion by staying away from each
other and thereby inducing spin polarization \cite{kubo} ). Later on Wigner
pointed out that correlation effects ( which also act between electrons with antiparallel spins ) destroy the possibility of ferromagnetism in a free electron
gas. In contrast, Stoner adopted a phenomenological approach and impressed an exchange field on free
electrons, and
discussed the possibility of ferromagnetism\footnote{Stoner approach is
analogous to Weiss approach in which Weiss incorporated mean molecular field on atomic moments thus
generalizing Langevin's theory of paramagnetism to ferromagnetism. }. The occurance of
ferromagnetism is given by the Stoner condition\[  I g (E_F)>1. \] Here I is the
Stoner exchange parameter and $g(E_F)$ is EDOS (Electronic Density of States) at the Fermi energy.
The origin of exchange field in Stoner model can be attributed to intra-atomic
Hund's mechanism \cite{kubo}.

 In the present paper we consider Stoner model for a ferromagnetic metal which is tuned near to its
magnetic instability $(Ig(E_F)\sim 1)$, and  which is placed in a
strong external magnetic field. EDOS  gets modified due to Landau level formation in the presence of external magnetic
field. We study the effect of  modified EDOS on the
Stoner condition. It turns out that under the action of strong magnetic field Stoner condition
becomes a function of external magnetic field through modified EDOS ($g(E_F,H)$).  As the strength
of
external magnetic field is changed the system oscillates between paramagnetic phase (when
$I g(E_F,H)<1 $) and ferromagnetic phase (when $I g(E_F,H)>1 $).
We also suggest an experiment to observe this effect.

The plan of the paper is as follows. In section (2) we present the derivation of Pauli susceptibility to set the stage for further development. In section (3) we study the effects of magnetic field on the EDOS and present a correction term to Pauli susceptibility. In section (4) we
present the oscillatory effects in Pauli susceptibility using Poisson summation formula. Then
generalizing this treatment we  presented  oscillatory effects in Stoner
susceptibility in section (5). We conclude that under the action of 
external magnetic field there are oscillations in EDOS as a function of external
magnetic field, which allows the system to
oscillate between paramagnetic phase and ferromagnetic phase alternatively through Stoner
condition.

\section{Pauli paramagnetic susceptibility}
Pauli paramagnetic susceptibility is due to the intrinsic angular momentum of the electrons.
Let $g(E)$ be the EDOS of free electrons with given polarization in zero magnetic field.
When an external magnetic field is applied electronic energy changes by $ E_k\pm\mu_BH .$
The total spin imbalance is given by
 

\beq \triangle N=\int_{0}^{\infty}{dE g(E) \left(f(E-\mu_B H)- f(E+\mu_B H)\right)}.
\eeq
Here $ f(E-\mu_BH)$ is the Fermi-Dirac distribution
function for electrons that are aligned along the field direction and $ f(E+\mu_BH) $ for those
electrons that are aligned in the opposite direction.
The induced magnetization is given by 
\beq M=\mu_B\int_0^{\infty} dE g(E) \left({(f(E-\mu_B H)-f(E +\mu_B H)}\right), \eeq
or 
 \beq M=2\mu_B^2 H\int_0^{\infty}{dE g(E)\left(\frac{f(E-\mu_B H)-f(E +\mu_B H)}{2\mu_B
H}\right)},\eeq

\beq M \simeq 2\mu_B^2 H\int_0^{\infty}{dE g(E)\left(-\frac{\partial f(E)}{\partial
E}\right)},\eeq
as  $\mu_BH << k_BT<< E_F $, we have
\beq \left(-\frac{\partial f(E)}{\partial E}\right)\simeq\delta(E-E_F),\eeq therefore 
equation (4) can be written as

\beq M=2\mu_B^2 H g(E_F),\eeq
or
\beq \chi_{Pauli}^\circ =2\mu_B^2 g(E_F),\eeq
as $ M=\chi_{Pauli}^\circ  H. $
This shows that Pauli susceptibility depends on density of states at Fermi level, and it is the
standard derivation\cite{prls}.

\section{Effect of magnetic field on the density of states and higher order term in the Pauli susceptibility}

The expression for the density of states per unit energy per unit volume in the case of free electrons is given by\cite{kit}
\beq g(E)= \frac{\sqrt{2E}m^{\frac{3}{2}}}{\pi^2\hbar^3}.\eeq

Under the action of an external magnetic field energy levels of a free electron gas split into Landau levels\cite{prls}\cite{snbrg}:
\beq E=(2n+1)\mu_BH +\frac{p_z^2}{2m}.\eeq
Since $p_z= \hbar k_z$ therefore
\beq k_z= \sqrt{\frac{2m}{\hbar^2}}(E-(2n+1)\mu_BH)^\frac{1}{2} \eeq
The number of electronic energy levels for z-direction is $k_z\left(\frac{L_z}{2\pi}\right)$. 
Including transverse plane degeneracy \cite{prls}, number of electronic energy levels up to energy
E is given by
\beq  N(E)=\frac{2\sqrt{2m}V e H}{(2\pi \hbar)^2c}\sum_n {(E-(2n+1)\mu_B H)^\frac{1}{2}}. \eeq
From the above equation EDOS is given by
\beq g(E,H)=\frac{dN(E)}{dE}=\frac{2\sqrt{2m}VeH}{(2\pi\hbar)^2 c}\frac{d}{dE}\left(\sum_n(E-(2n+1)\mu_{B}H)^\frac{1}{2}\right). \eeq 
Before differentiating, let us first simplify the summation. 
\beq S= \sum_n {(E-2n\mu_B H-\mu_B H)^\frac{1}{2}}=(2\mu_B H)^\frac{1}{2}\sum_n\left(\frac{E}{2\mu_B H}-n-\frac{1}{2}\right)^\frac{1}{2}.\eeq
Define $\epsilon=\frac{E}{2\mu_BH}$ and write S as $(2\mu_BH)^\frac{1}{2}f(\epsilon)$, where
$$f(\epsilon)= \sum_{n}\left(\epsilon-n-\frac{1}{2}\right)^\frac{1}{2}.$$
From the Poisson summation formula (Appendix A), and  under the condition\footnote{For more information see appendix B} when $\mu_BH <<k_BT<< E_F $,  
we have
\beq \sum_n \left(\epsilon-n-\frac{1}{2}\right)^\frac{1}{2}\simeq\frac{2}{3}\epsilon^{\frac{3}{2}}-\frac{1}{48}\epsilon^{-\frac{1}{2}}.\eeq
Under this approximation, EDOS takes the form:

\beq g(E,H) \simeq \frac{2\sqrt{2m}VeH}{(2\pi\hbar)^2
c}\left(\frac{\sqrt{E}}{\mu_{B}H}+\frac{1}{24}\left(\frac{\mu_{B}H}{E^{\frac{3}{2}}}\right)\right)
, \eeq
\beq g(E,H)\simeq\frac{\sqrt{2}m^\frac{3}{2}}{\pi^2
\hbar^3}\left(\sqrt{E}+\frac{1}{24}\left(\frac{(\mu_B
H)^2}{E^{\frac{3}{2}}}\right)\right). \eeq

As  $\chi_{Pauli}=2\mu_B^2g(E_F,H)$ therefore we get
\beq \chi_{Pauli}(H)\simeq\chi_{Pauli}^\circ \left(1
+\frac{1}{24}\left(\frac{\mu_BH}{E_F}\right)^2\right), \eeq
where $\chi_{Pauli}^\circ=2\mu_B^2g(E_F).$
Here the correction term $\left( \frac{\mu_BH}{E_F} \right)^2 $ in the Pauli susceptibility originates from the deformed EDOS
due to external magnetic field. It is small correction as $\mu_BH << E_F$.

\section{Oscillatory effects in Pauli susceptibility}
The expression for the EDOS at low temperature i.e. $(k_BT\lesssim\mu_BH<< E_F)$ under the action of an
external magnetic field is given by (after differentiating eqn.(11) w.r.t. E)

\beq
g(E,H)=\frac{2\sqrt{2m}VeH}{(2\pi\hbar)^2c}\left(\frac{1}{2}\sum_n(E-(2n+1)\mu_{B}H)^{-\frac{1}{
2 }}\right).   \eeq
 Define $\lambda=\frac{2\sqrt{2m}VeH}{(2\pi\hbar)^2 c},$ with this the  
above equation takes the form:
\beq g(E,H)=\lambda \left(\frac{1}{2}\sum_n(E-(2n+1)\mu_{B}H)^{-\frac{1}{2}}\right).   \eeq
As $\chi_{Pauli}(H) = 2\mu_B^2g(E_F,H),$ therefore the expression for Pauli susceptibility is given
by

\beq \chi_{Pauli}(H)=\mu_B^2 \lambda\ \sum_n\left(E_F-(2n+1)\mu_B H\right)^{-\frac{1}{2}} \eeq
or
\beq \chi_{Pauli}(H)=\mu_B^2 \lambda\sum_n(2\mu_B H)^{-\frac{1}{2}}\left(\frac{E_F}{(2\mu_B
H)}-n-\frac{1}{2}\right)^{-\frac{1}{2}}. \eeq
Define $ \tilde{\epsilon}_F=\frac{E_F}{(2\mu_B H)}$. As $\mu_BH<< E_F $ therefore $ \tilde{\epsilon}_F >> 1$. The above equation can be written as:
\beq \chi_{Pauli}(H)=\frac{\mu_B^2 \lambda}{(2\mu_B
H)^{\frac{1}{2}}}\sum_n\left(\tilde{\epsilon}_F-n-\frac{1}{2}\right)^{-\frac{1}{2}}. \eeq

As $\tilde{\epsilon}_F>>1,$ the upper limit can be restricted to $\tilde{\epsilon}_F$ instead
of $(\tilde{\epsilon}_F-\frac{1}{2}) $. The summation can be written as
\beq  S=\sum_{n=0}^{\tilde{\epsilon}_F}{(\tilde{\epsilon}_F
-n)^{-\frac{1}{2}}}=\lim_{\delta \to \ 0}\int_{-\delta}^{\tilde{\epsilon}_F+\delta}{dx\sum_{n=0}^{
\tilde{\epsilon}_F}{(\tilde{\epsilon}_F-x)^{-\frac{1}{2}}}\delta(x-n)}, \eeq
on writing the delta function as Fourier sum\footnote{Here we will do the exact treatment i.e., including the oscillatory terms as we have $ k_BT\lesssim \mu_BH<<E_F $. Refer Appendix B for details.}, we obtain 
\beq S = \lim_{\delta \to \ 0}\int_{-\delta}^{
\tilde{\epsilon}_F+\delta}{dx{
(\tilde{\epsilon}_F-x)}^{-\frac{1}{2}}}\sum_{k=-\infty}^{+\infty}{e^{2\pi i k x}} ,\eeq
therefore equation (22) can be written as


\beq \chi_{Pauli}(H)=\frac{\mu_B^2 \lambda}{(2\mu_B H)^{\frac{1}{2}}}
\sum_{k=-\infty}^{+\infty}\lim_{\delta\to\ 0}\int_{-\delta}^{\tilde{\epsilon}_F+\delta}{dx{
(\tilde{\epsilon}_F-x)}^{-\frac{1}{2}}}{e^{2\pi i k x}}. \eeq
Before integrating let us first simplify\footnote{Considering only real part as static susceptibility is a real quantity.}  $\sum_{k=-\infty}^{+\infty}{e^{2\pi i k x}}.$ We have

\beq\sum_{k=-\infty}^{+\infty}{e^{2\pi i k x}}=\sum_{k=-\infty}^{+\infty}{\left(cos(2\pi kx)+isin(2\pi kx)\right)}=\left(1+2\sum_{k=1}^{\infty}cos2\pi kx\right). \eeq
By using this result the equation (25) takes the form:
\beq \chi_{Pauli}(H)=\frac{\mu_B^2 \lambda}{(2\mu_B H)^{\frac{1}{2}}}
\lim_{\delta\to\ 0}\int_{-\delta}^{\tilde{\epsilon}_F+\delta}{dx
{(\tilde{\epsilon}_F-x)}^{-\frac{1}{2}}}{\left(1+2\sum_{k=1}^{\infty}cos2\pi kx\right)}, \eeq


\beq \chi_{Pauli}(H)=\frac{2\mu_B^2 \lambda}{(2\mu_B H)^{\frac{1}{2}}}\sqrt{\tilde{\epsilon}_F}
+\frac{2\mu_B^2 \lambda}{(2\mu_B H)^{\frac{1}{2}}}
\sum_{k=1}^{\infty}\int_{0}^{\tilde{\epsilon}_F}{dx{(\tilde{\epsilon}_F-x)}^{-\frac{1}{2}}}\cos2\pi kx
.\eeq
Let us simplify the integral $
I=\int_{0}^{\tilde{\epsilon}_F}{dx{(\tilde{\epsilon}_F-x)}^{-\frac{1}{2}}}\cos2\pi kx ,$

\beq I= \int_{0}^{\tilde{\epsilon}_F}{dx{(\tilde{\epsilon}_F-x)}^{-\frac{1}{2}}}\cos2\pi kx
=\frac{1}{\sqrt{k}} \left(cos(2\pi\tilde{\epsilon}_F
k)F_C(\sqrt{4k\tilde{\epsilon}_F})+sin(2\pi\tilde{\epsilon}_F
k)F_S(\sqrt{4k\tilde{\epsilon}_F})\right).\eeq
Here the $F_C(\sqrt{4k\tilde{\epsilon}_F}) $ and  $F_S(\sqrt{4k\tilde{\epsilon}_F}) $ are the Fresnel
integrals,
 therefore the expression for
$\chi_{Pauli}(H)$ is given by
\beq
\chi_{Pauli}(H)=\chi_{Pauli}^\circ\left(1+\sqrt{\frac{2\mu_BH}{E_F}}\sum_{k=1}^{\infty}{\frac{
(cos(k\pi\frac{ E_F }{\mu_BH})F_C(\sqrt{2k\frac{ E_F}{\mu_BH}})+sin( k\pi \frac{ E_F
}{\mu_BH})F_S(\sqrt{2k\frac{ E_F}{\mu_BH}}))}{\sqrt{k}}}\right) . \eeq
This expression is  Lifshitz-Kosevich formula but including only the spin part.
 Let us define
\beq f\left(\frac{E_F}{\mu_BH}\right)= \sqrt{\frac{2\mu_BH}{E_F}}\sum_{k=1}^{\infty}{\frac{\left(
		cos(k\pi\frac{ E_F }{\mu_BH})F_C(\sqrt{2k\frac{ E_F}{\mu_BH}})+sin(k\pi\frac{ E_F
		}{\mu_BH})F_S(\sqrt{2k\frac{ E_F}{\mu_BH}})\right)}{\sqrt{k}}}.    \eeq
When the $k_BT\lesssim \mu_BH<< E_F$ then the expression for Pauli susceptibility is given by 

\beq \chi_{Pauli}(H)=\chi_{Pauli}^\circ\left(1+f\left(\frac{E_F}{\mu_BH}\right)\right),  \eeq
which is  an oscillatory function.

\section{Oscillatory effects in Stoner susceptibility}
 From  the expression for Stoner susceptibility \cite{kubo}  we have
\beq \chi_{Stoner} = \frac{\chi_{Pauli} }{(1- I \chi_{Pauli} )} ,\eeq
where I is an effective exchange energy.



Substituting the expression for  $\chi_{Pauli}$ from equation (32), 
the expression for Stoner susceptibility is given by

\beq \chi_{Stoner}=\frac{\chi_{Pauli}^\circ }{1- f(\frac{E_F}{\mu_B H})-I \chi_{Pauli}^\circ },\eeq
because $f\left( \frac{E_F}{\mu_BH}\right) << 1 $ we divided numerator and denominator by $\left(  1+f\left( \frac{E_F}{\mu_BH}\right)\right) $. Set  $\mu_B =1 $ for simplicity.

 Now the Stoner condition $I g_\circ (E_F)>1 $ changes to $ f\left(\frac{E_F}{H}\right) +
Ig_\circ (E_F) > 1$ ( $g_\circ(E_F) $ is the total EDOS including both spin directions).
Let us take a special case where the system is near to Stoner instability $I g_\circ (E_F)\simeq  1 ,$ we get

$$ f\left( \frac{E_F}{H}\right) > 0, $$ 
for the condition of ferromagnetism when the metal is near its magnetic instability. But it depends upon external magnetic field
strength H and it is an oscillatorty function (figure 1) of $\eta=\frac{E_F}{H}$ and it changes sign also.
Thus when magnetic field H  is varied, system oscillates between paramagnetic and ferromagnetic phases.

\begin{figure}[!ht]
\begin{center}
\includegraphics[height=5.5cm]{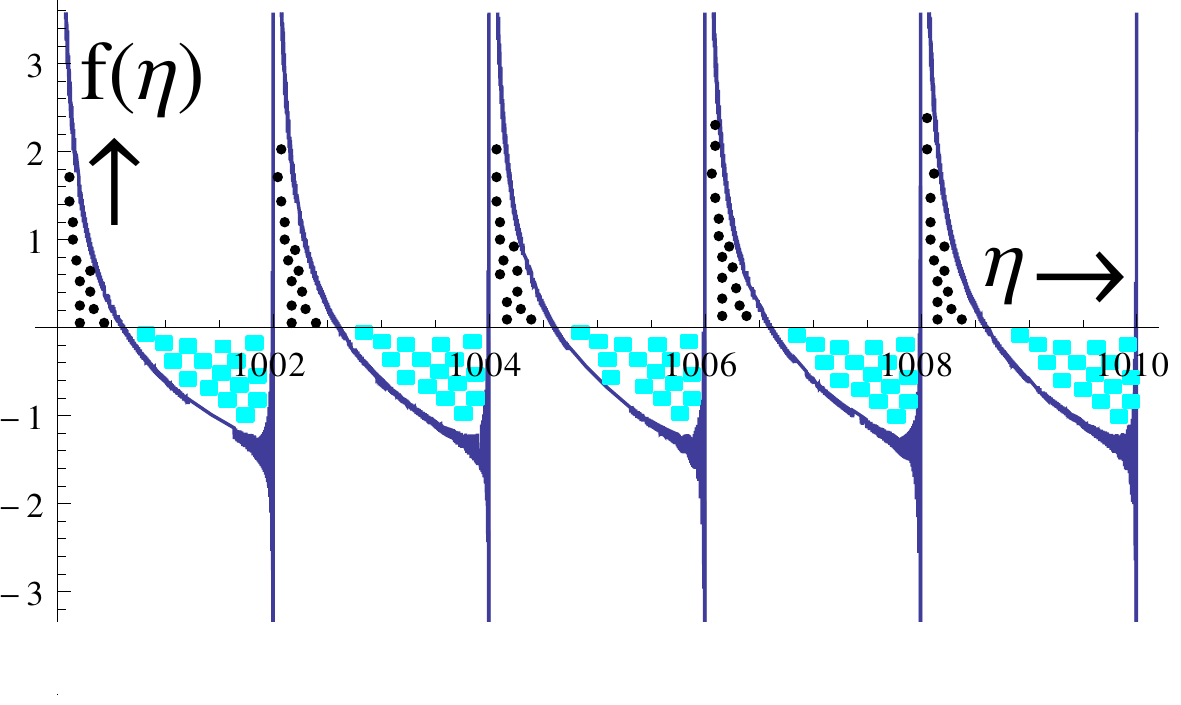}
\caption{Oscillatory behaviour of function $f(\eta=\frac{E_F}{H})$ as a function of external
magnetic
field strength. Here  reactangles shows paramagnetic regions below the $\eta$ axis and  dots above the $\eta$ axis shows ferromagnetic
regions. When magnetic field is varied i.e.  when $\eta$ is varied the system oscillates between ferromagnetic regions and paramagnetic regions.}
\end{center}
\end{figure}

\section{Discussion and conclusion}
The above figure 1 shows that there are oscillations in Stoner susceptibility as a function of
external magnetic field strength (H).  An interesting behaviour is noticed when $f(\eta) > 0$, the
system lies in ferromagnetic phase and when $f(\eta)<0 $ the behaviour of system switches to
paramagnetic phase. This kind of oscillatory behaviour shows that the system under the influence of
strong external magnetic field oscillates between paramagnetic phase and ferromagnetic phase as a
function of H. The oscillatory behaviour arises from the deformed EDOS of the system in the
presence of strong external magnetic field.

One can observe such oscillatory behaviour
experimentally with a special experimental set-up as shown in figure 2. The sample ( for example the material  $HfZr_2$ which can be tuned near to its Stoner instability \cite{ir})  is placed inside a coil and then strong external magnetic field is applied on the
sample which is along the axis of the coil.  
When magnitude of field is varied the sample's magnetization oscillates as depicted in figure 1. Due to the changing 
magnetization of sample an induced $ e.m.f$  is produced in the coil which can be amplified if needed and the output can be given to an oscilloscope.

\begin{figure}[!ht]
\begin{center}
\includegraphics[height=10cm]{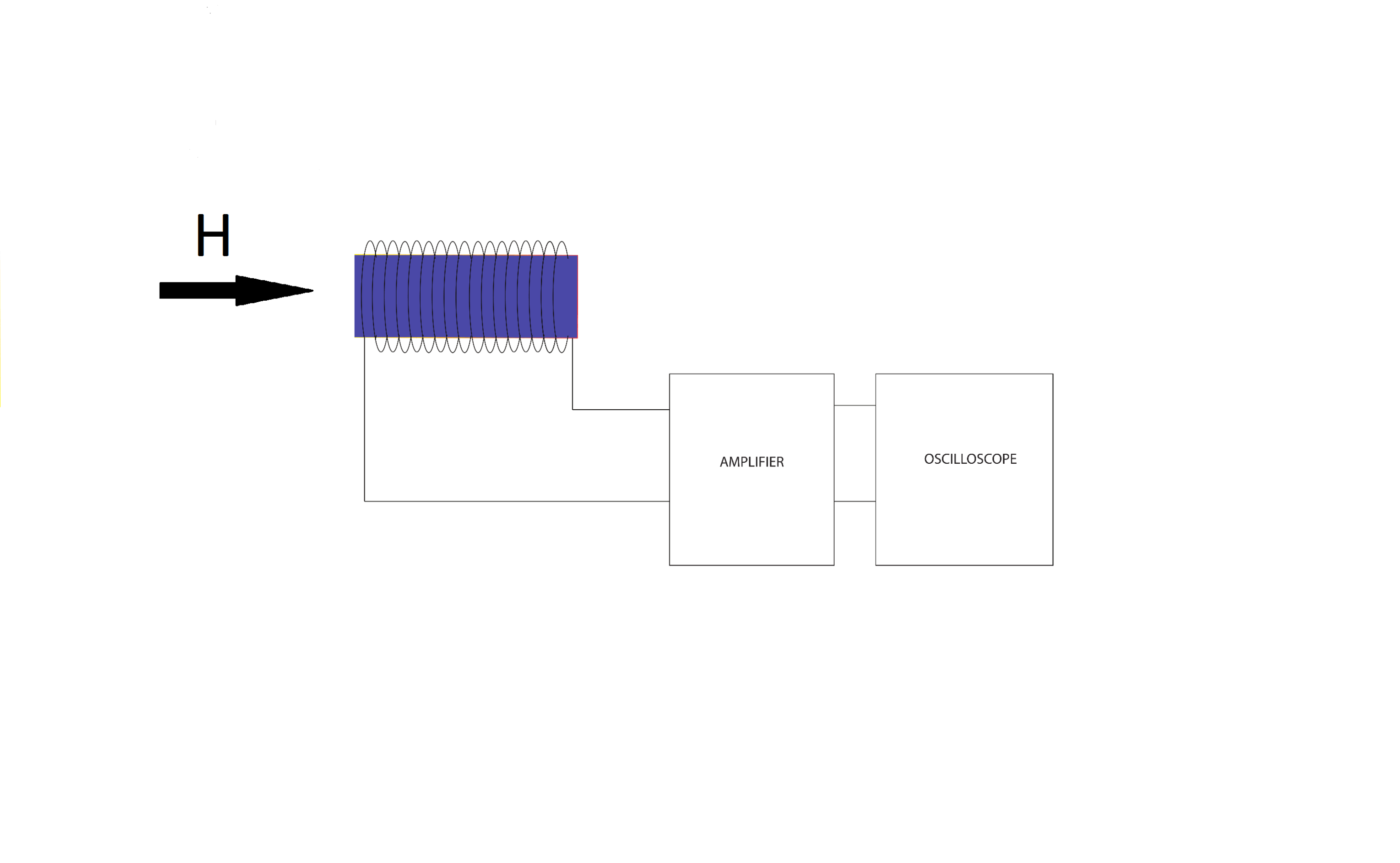}
\caption{Experimental set-up for the study of oscillatory behaviour.}
\end{center}
\end{figure}

We also suggest an important application of this effect. In the current topic of anomalous transport and thermal properties of magnetic material tuned near to their quantum criticality it is important to tune and control the system near and around a QCP (Quantum Critical Point). Various methods like doping, pressure, and magnetic field is used for this purpose \cite{rfrns}\cite{rf1}\cite{rf2}\cite{rf3} our method could be a new addition to such methods. The present method will lead to H-T phase diagrams of the form depicted in figure 3. The alternative phases result due to oscillations in $f(\eta)$ as explained in figure 1. At high temperature oscillations will vanish and regions in figure 3 will have dome like structures.
\begin{figure}[!ht]
\begin{center}
\includegraphics[height=8cm]{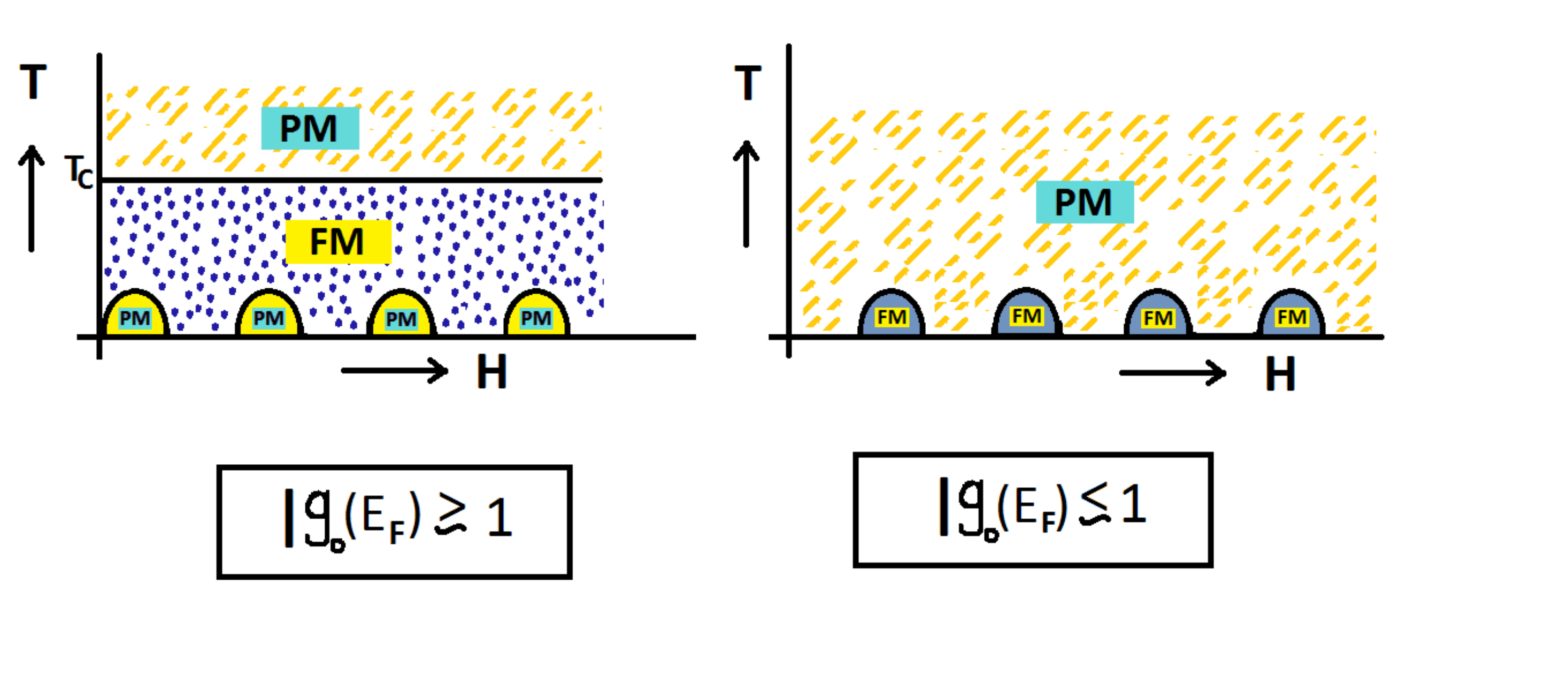}	
\caption{H-T phase diagrams. Here PM stands for Paramagnetism and FM stands for Ferromagnetism.}
\end{center}
	
\end{figure}

\section{Appendix A}
We will prove equation (14) by two methods. In the first method we use Poisson summation formula and in the second method we use Euler-Maclaurin sum formula.

\section*{Proof of equation (14) by using Poisson summation formula }
 Starting from $$ f(\epsilon)= \sum_n\left(\epsilon-n-\frac{1}{2} \right)^\frac{1}{2}, $$ as
$\epsilon >>1$, the upper limit of sum over $ n $ can be restricted to $n\leq\epsilon $ instead of $n\leq (\epsilon-\frac{1}{2})$
therefore the $f(\epsilon)$ is given by
\beq f(\epsilon)\simeq  \sum_n\left(\epsilon-n \right)^\frac{1}{2}. \eeq
From Poisson summation formula \cite{prls} 
\beq \sum_{n=0}^{\infty} F(n)=\sum_{l=-\infty}^{+\infty}{(-1)^l \int_{0}^{\infty}F(x) e^{2\pi
i lx} dx }. \eeq
Summation in equation (35) can be evaluated:
$$ F(n)=\sqrt{\epsilon-n}, $$ 
\beq \sum_{n=0}^{\epsilon}\left(\epsilon-n\right)^\frac{1}{2} \simeq \sum_{l=-\infty}^{+\infty}{(-1)^l
\int_{0}^{\epsilon}(\epsilon-x)^\frac{1}{2} e^{2\pi i lx} dx}.\eeq
On integrating by parts we get

\beq \sum_{n=0}^{\epsilon}\left(\epsilon-n\right)^\frac{1}{2}\simeq
\int_{0}^{\epsilon}{dx (\epsilon-x)^\frac{1}{2}} + \sum_{l=-\infty ,l\ne 0}^{+\infty}{\frac{(-1)^{l+1}
}{l} \frac{\epsilon^\frac{1}{2}}{2\pi i}}+ \sum_{l=-\infty , l\ne 0}^{+\infty}{\frac{(-1)^{l} }{l^2}
\frac{\epsilon^{-\frac{1}{2}}}{8\pi^2 }} + (oscillatory ~ terms). \eeq
The second term in R.H.S. of above equation converges to zero\cite{wlf} ( as each positive term cancel with its symmetric negative term) and $\sum_{-\infty}^{+\infty}{\frac{(-1)^l}{l^2}}= -\frac{\pi^2}{6}$, therefore we have
\beq \sum_n\left(\epsilon-n\right)^\frac{1}{2}\simeq
\frac{2}{3}\epsilon^\frac{3}{2}-\frac{1}{48}\epsilon^{-\frac{1}{2}}.\eeq
We explian the issue of oscillatory terms in Appendix B.
 
 \section*{Proof of eqn. (14) by using Euler-Maclaurin sum formula }
 Starting from \beq f(\epsilon)= \sum_n \left(\epsilon-n-\frac{1}{2}\right)^\frac{1}{2}.\eeq
 From the Euler-Maclaurin formula\cite{ldem} we have
 \beq \sum_{n=0}^{\infty}g(n+\frac{1}{2})\simeq \int_{0}^{\infty}g(x)dx + \frac{1}{24}g^\p
 (x)|_{x=0}. \eeq
 As $g\left( n+\frac{1}{2}\right)=\left(\epsilon-n-\frac{1}{2}\right)^\frac{1}{2}$,
 
 \beq
 \sum_{n=0}^{\epsilon}\left(\epsilon-n-\frac{1}{2}\right)^\frac{1}{2}\simeq \int_{0}^{\epsilon}{
 	(\epsilon-x)^\frac{1}{2}}dx-\frac{1}{48}(\epsilon-x)^{-\frac{1}{2}}|_{x=0} ,\eeq 
as $\epsilon \to\infty $ we can use (41)
 \beq
 \sum_{n=0}^{\infty}\left(\epsilon-n-\frac{1}{2}\right)^{\frac{1}{2}}\simeq \frac{2}{3}\epsilon^{
 	\frac {
 		3}{2}}-\frac{1}{48}\epsilon^{-\frac{1}{2}} .\eeq

\section*{Appendix B}
\section*{When are oscillations in susceptibility  important and when not ?}

\begin{figure}[!ht]
	\begin{center}
		\includegraphics[height=5cm]{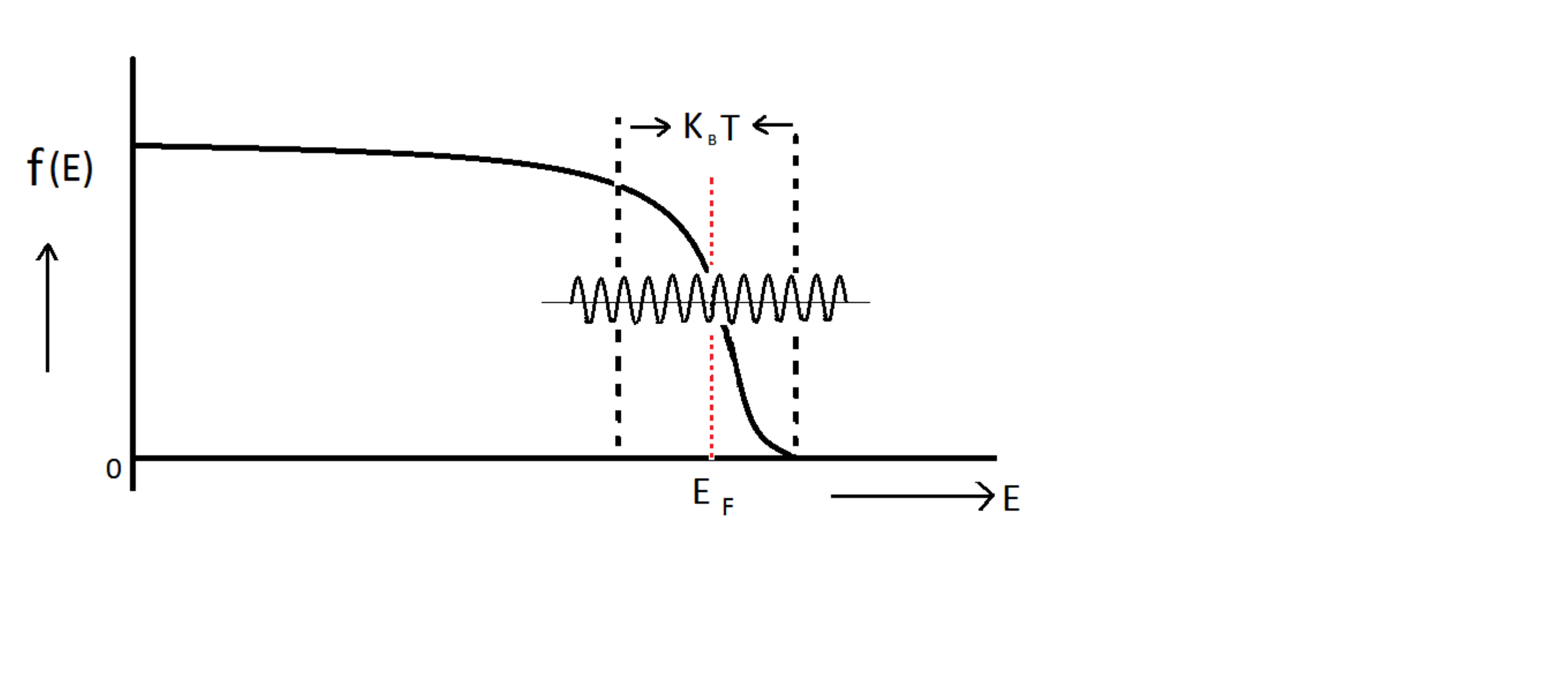}
		\caption{When $\mu_BH<<k_BT$}
	\end{center}

\end{figure}

Case 1. Oscillations are not important:  when $ \mu_BH << k_BT << E_F $ there are large number of oscillations over an energy scale of $k_BT $ (figure 4). These oscillating terms give no contribution in the Pauli susceptibility due to thermal averaging effect. This can be appreciated in the following way. The oscillatory term appears in a more careful treatment of the summation (14). The sum (from equation 38) can be exactly written as 
\beq \sum_{n=0}^{\epsilon}(\epsilon-n)^{\frac{1}{2}}=\int_{0}^{\epsilon}(\epsilon-x)^{\frac{1}{2}}+\sum_{l=-\infty}^{+\infty}\left({\frac{(-1)^{l+1}}{l}\frac{\epsilon^{\frac{1}{2}}}{2\pi i}}+{\frac{(-1)^l}{l^2}\frac{\epsilon^\frac{1}{2}}{8\pi^2}}+\underbrace{\frac{1}{16\pi^2}{\frac{(-1)^l}{l^2}\int_{0}^{\epsilon}{(\epsilon-x)^{-\frac{3}{2}}} e^{2\pi i lx}dx}}_{Oscillatory ~ term}\right).  
\eeq
The last term on the RHS of the above equation is the oscillatory term. To understand when it is important and when it is not important consider equation (4) and substitute the expression for $g(E,H)$ from equation (19) into equation (4):
\beq \chi_{Pauli}\simeq\mu_B^2 \lambda\int_{0}^{\infty}{dE\left(-\frac{\partial f(E)}{\partial E} \right) \sum_n(E-(2n+1)\mu_BH)^{-\frac{1}{2}}}.
\eeq 
Now the summation includes the oscillatory term in the above equation. The derivative $\left(-\frac{\partial f(E)}{\partial E} \right)$ in the above equation is not exactly a delta function, but it is a Gaussian function of width $k_BT$ centered around $E_F$. If there are many oscillations over $k_BT$ then they average out in the integral over E in the above equation. This leads to omitting of oscillations.

 Case 2. Oscillations are important: when $k_BT\lesssim \mu_BH<< E_F $, in this case  the temperature is low and $ \mu_BH $ is the same order as  $k_BT $ then the contribution of oscillations is to be taken into account (figure 5). The treatment given in section (4) takes this into account exactly.

\begin{figure}[!ht]
	\begin{center}
		\includegraphics[height=5cm]{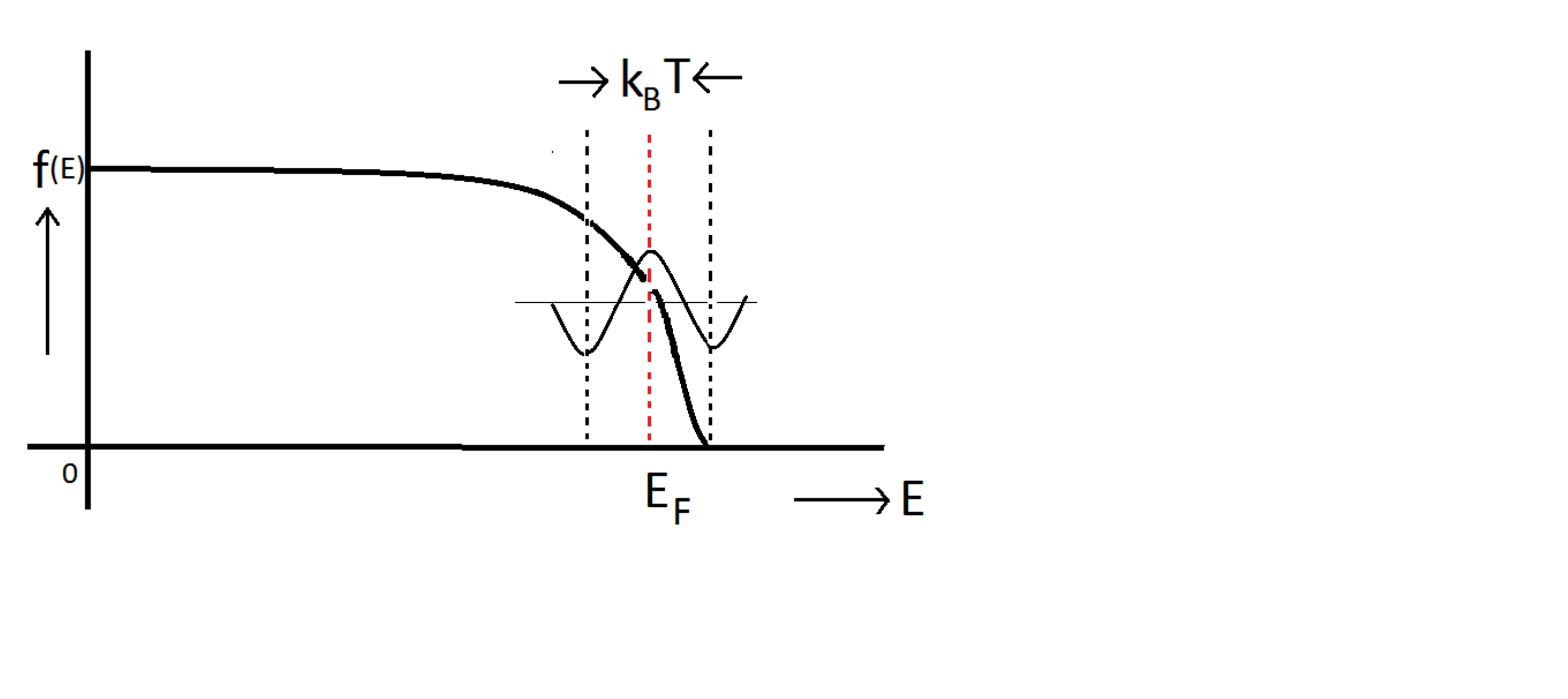}
		\caption{ When $ k_BT\lesssim \mu_BH$}
	\end{center}

\end{figure}

\end{document}